\documentclass{emulateapj}
\usepackage{graphicx}

\slugcomment{To appear in ApJ}
\shorttitle{Black hole spin orientations and magnitudes}
\shortauthors{Dotti et al.}

\def\msun{\rm M_{\odot}}
\def\mbh{M_{\rm BH}}
\def\jbh{{\mathbf J}_{\rm BH}}
\def\jbhdir{{\hat {\mathbf J}_{\rm BH}}}
\def\jdisk{{\mathbf J_{\rm disk}}}
\def\jdiskdir{{\hat {\mathbf J}_{\rm disk}}}
\def\fedd{f_{\rm Edd}}

\def\ldir{{\mathbf {\hat l}}}
\def\jtotal{{\mathbf J}_{\rm tot}}
\def\thetabh{{\theta_{\rm BH}}}
\def\bfl{{\mathbf L}}
\def\bflhat{{\mathbf {\hat l}}}

\def\simlt{\mathrel{\rlap{\lower 3pt\hbox{$\sim$}}\raise 2.0pt\hbox{$<$}}}

\def\simgt{\mathrel{\rlap{\lower 3pt\hbox{$\sim$}} \raise 2.0pt\hbox{$>$}}}

\def\lsim{\mathrel{\rlap{\lower 3pt\hbox{$\sim$}}\raise 2.0pt\hbox{$<$}}}

\def\gsim{\mathrel{\rlap{\lower 3pt\hbox{$\sim$}} \raise 2.0pt\hbox{$>$}}}

\def\msunpc3{\msun~{\rm {pc^{-3}}}}

\newcommand{\be}{\begin{equation}}

\newcommand{\ee}{\end{equation}}

\begin{document}

\title{
On the orientation and magnitude of the black hole spin in galactic nuclei
}

\author{
M. Dotti\altaffilmark{1,2},  M. Colpi\altaffilmark{1,2}, S. Pallini\altaffilmark{1},
  A. Perego\altaffilmark{3}, M. Volonteri\altaffilmark{4,5}
%
} \altaffiltext{1}{Dipartimento di Fisica G. Occhialini, Universit\`a
  degli Studi di Milano Bicocca, Piazza della Scienza 3, 20126 Milano,
  Italy E-mail: {\sf Massimo.Dotti@mib.infn.it}}
\altaffiltext{2}{INFN, Sezione di Milano-Bicocca, Piazza della Scienza 3, 20126 Milano, Italy} 
\altaffiltext{3}{Department of Physics,
  University of Basel, Klingerbergstr. 82, 4056 Basel, Switzerland}
\altaffiltext{4}{Astronomy Department, University of Michigan,
  Ann Arbor 48109, USA}
\altaffiltext{5}{Institut d’Astrophysique de Paris, 98
  bis Bd Arago, Paris, 75014, France}


\begin{abstract}
Massive black holes in galactic nuclei vary their mass $\mbh$ and spin
vector $\jbh$ due to accretion.  In this study we relax, for the first
time, the assumption that accretion can be either chaotic, i.e.  when
the accretion episodes are randomly and isotropically oriented, or
coherent, i.e. when they occur all in a preferred plane.  Instead, we
consider different degrees of anisotropy in the fueling, never
confining to accretion events on a fixed direction.  We follow the
black hole growth evolving contemporarily mass, spin modulus $a$ and
spin direction.  We discover the occurrence of two regimes. An early
phase ($\mbh\lsim 10^7\,\msun$) in which rapid alignment of the black
hole spin direction to the disk angular momentum in each single
episode leads to erratic changes in the black hole spin orientation
and at the same time to large spins ($a\sim 0.8$).  A second phase
starts when the black hole mass increases above $\gsim 10^7\,\msun$
and the accretion disks carry less mass and angular momentum
relatively to the hole.  In the absence of a preferential direction
the black holes tend to spin-down in this phase.  However, when a
modest degree of anisotropy in the fueling process (still far from
being coherent) is present, the black hole spin can increase up to $a
\sim 1$ for very massive black holes ($\mbh\gsim 10^8\,\msun$), and
its direction is stable over the many accretion cycles.  We discuss
the implications that our results have in the realm of the
observations of black hole spin and jet orientations.
\end{abstract}

\section{Introduction}

Massive black holes (BHs) that inhabit in the nuclei of massive
galaxies are described by only two parameters: their mass $\mbh$ and
spin $\jbh$.

 Tens of BH masses in quiescent, nearby galaxies have been directly
 estimated to date studying the dynamics of stars and gas and these
 masses all fall within $\gsim 10^6 \,\msun - 10^{10}\, \msun$
 (e.g. G\"ultekin et al. 2009, McConnell et al. 2011 and references
 therein).  Similarly, tens of thousands BH masses in active
   nuclei have also been estimated using empirical relations, and,
   while for quasars their median mass is in excess of $10^8\,\msun$
   (e.g., Vestergaard et al. 2008), faint AGN are powered by BHs with
   masses as small as $\sim 10^5\,\msun$ (Peterson et al. 2005).

Measurements of the dimensionless spin parameter $a$ are more
controversial.  The parameter $a$ has been recently measured only for
few AGNs (e.g. Brenneman \& Reynolds 2006; Schmoll et al. 2009; de la
Calle Perez et al. 2010; Patrick et al. 2011a,b; Gallo et al. 2011;
Brenneman et al. 2011) through X-ray spectroscopy. However, several
free parameters enter in the estimate, creating a severe degeneracy
problem.  For example, different groups obtained different esitmates
of $a$ in MCG-6-30-15: $a > 0.98$ (Brenneman \& Reynolds 2006), $a =
0.86 \pm 0.01$ (de la Calle Perez et al. 2010) and $a =
0.49^{+0.20}_{-0.12}$ (Patrick et al. 2011a). Even more problematic is
the case of NGC 3783, for which different groups found $a > 0.88$
(Brenneman et al. 2011) and $a < 0.32$ (Patrick et al. 2011b).
Measuring and understanding spins is crucial to assess the cosmic
evolution of massive BHs.  Firstly, spins affect the
accretion-luminosity conversion efficiency; highly spinning BHs can
convert up to $\sim 40\%$ of the accreted matter into radiation,
making them more luminous, albeit making their growth slowlier.
Secondly, the spin paradigm assumes that radio jets observed in AGNs
are launched by highly spinning BHs (Blandford \& Znajek
1977). Lastly, spins dramatically affect the gravitational recoil
suffered by the remnant BH after a binary merger. It has in fact been
shown that highly spinning BHs can experience kicks up to 5000 km
s$^{-1}$ depending on their progenitor spin magnitude and orientation
(Campanelli et al. 2007; Baker et al. 2008; Herrmann et al. 2007;
Schnittman \& Buonanno 2007; Lousto \& Zlochower 2011; Lousto et
al. 2012).  These super-kicks are sufficient to eject the remnant from
the deepest potential well of the most massive galaxies, with
potentially important implications for the occupation fraction of
massive BHs in galaxies (Schnittman 2007; Volonteri 2007; Volonteri,
Haardt \& G\"ultekin 2008; Volonteri, G\"ultekin \& Dotti 2010).

BH masses and spins are thought to build up through gas accretion and
BH mergers, and their history of growth erases the initial values of
$M_{\rm BH}$ and $a$.  The main driver of the BH spin evolution is gas
accretion (Berti \& Volonteri 2008; Fanidakis et al. 2011; Barausse
2012)\footnote{With the possible exception of the most massive BHs in
  massive, gas-poor, low-redshift ellipticals (e.g. Fanidakis et
  al. 2011).}. Two different accretion scenarios have been proposed to
date: (i) {\it coherent accretion}, in which the BHs accrete gas with
a well defined, almost constant, angular momentum direction, and (ii)
{\it chaotic accretion}, in which  parcels of gas accrete on the
BHs in randomly oriented planes (e.g. King et al. 2005; King \&
Pringle 2006; 2007).  The two models result in different BH evolutions
and different expected distributions of the spin magnitudes. Coherent
accretion keeps on adding angular momentum to the BHs in the same
direction and results in very high spins ($0.8 \lsim a \lsim 1$)
aligned with the angular momentum of the accreting material
(e.g. Dotti et al. 2010 and references therein). The chaotic case is
more subtle: gas accretion on a rotating BH on a retrograde orbit has
a larger last stable orbit than gas with angular momentum aligned with
the BH spin. As a consequence, retrograde accreting gas transfers to
the BH more (and negative) angular momentum per unit of mass than the prograde. If
retrograde and prograde accretion are equally probable (as implicitly
assumed in the chaotic scenario), the BH spin $a$ is biased toward low
values ($0 \lsim a \lsim 0.2$, King, Pringle \& Hofmann 2008; Berti \&
Volonteri 2008).

These two models cover the extreme cases in which the gas either flows
from a stable, fixed direction or from fully random directions.  In
this investigation we relax these extreme and unrealistic assumptions,
exploring the evolution of the BH mass and spin vector varying the
degree of anisotropy in the fueling process to mimic anisotropies
present in the gas in the nuclear regions of active galaxies. We
demonstrate how the evolution of the BH spin direction and magnitude
is coupled, with the evolution of the spin direction determining the
growth or reduction of $a$.

The paper is organized as follows: in Section 2 we describe the gas
dynamics and introduce the equations of evolution for the mass and
spin of the accreting BHs; in Section 3 we discuss our results; in
Section 4 we summarize the results and in Section 5 we discuss the
potential implications of our study on the growth of BHs and on jet
formation.

\section{Methodology}

We follow individual BH histories by tracing the evolution of the
black hole mass $\mbh$ and angular momentum vector ${\mathbf J}_{\rm
  BH}=(aGM_{\rm BH}^2/c)\hat{{\mathbf J}}_{\rm BH}$, where we denote
with $0 \leq a \leq 1$ the dimensionless spin parameter and with
$\hat{\mathbf J}_{\rm BH}$ the spin orientation.  A single history is
a sequence of multiple accretion episodes during which $\mbh$, $a$ and
$\hat{\mathbf J}_{\rm BH}$ vary upon time $t$ according to the recipes
described below.

\subsection{The accretion disk properties}

In every single accretion episode the BH is assumed to be surrounded
by a stationary geometrically thin, optically thick $\alpha-$disk
(Shakura \& Sunyaev 1973).  The accretion rate $\dot M$ is expressed
in terms of the Eddington factor $\fedd$, and the accretion efficiency
$\eta$ (which is a function of $a$, Bardeen 1970; Bardeen et
al. 1972) according to the relation
\begin {equation}
{\dot M}={\fedd \over \eta}{\mbh\over {\tau_{\rm S}}}=2\times 10^{-2} M_{\rm BH,6} \left(\frac{f_{\rm Edd}}{\eta_{0.1}}\right)\,\msun \, \rm yr^{-1},
\end{equation}
where $\eta_{0.1}$ is the efficiency in units of 0.1, $M_{\rm BH, 6}$
the BH mass in units of $10^6\,\msun$, ${\tau_{\rm S}}=\sigma_{\rm
  T}c/(4\pi Gm_{\rm p})=4.5\times 10^8$ yr, $\sigma_{\rm
  T}$ is the Thomson cross-section, and $m_{\rm p}$ is the proton mass.

We set the total mass of the disk $m_{\rm disk}$ at each accretion
episode equal to the minimum between $(i) \,m_{\rm cloud}$, i.e.  the
mass of a gas cloud that we assume to be available in a single feeding
episode, and $(ii)$ the self-gravitating mass $m_{\rm sg}$, the
largest for an $\alpha-$disk to be stable against fragmentation by its
own self gravity (e.g. Kolykhalov \& Sunyaev 1980).  The mass $m_{\rm
  cloud}$ is taken to be a constant, i.e. independent of the BH mass,
and carries two possible values $10^4\msun$ and $10^5 \msun.$ Within a
given accretion history, i.e. over the whole BH's life, the mass of
the cloud is kept fixed at either $10^4\msun$ and $10^5 \msun.$ The
self-gravitating mass $m_{\rm sg},$ for the $\alpha-$disk, is 
computed from the distance $R_{\rm disk,sg}$ at which the Toomre parameter
$Q\sim c_{\rm s}\Omega/\pi G \Sigma\lsim 1$, so that at radii
$R<R_{\rm disk,sg}$ the disk is stable (where $c_{\rm s}$ is the
central sound speed, $\Omega$ the Keplerian angular velocity of fluid
elements in the disk, and $\Sigma$ the surface mass density). This
yields
\begin{equation}\label{eqn:Rdisksg}
{R_{\rm disk,sg}\over R_{\rm G}}\approx  10^5 \alpha_{0.1}^{28/45}M_{\rm BH,6} ^{-52/45}\left ( {\fedd\over \eta_{0.1}}\right )^{-22/45} 
\end{equation}
where $\alpha_{0.1}=\alpha/0.1$ is the radial shear viscosity
parameter in units of 0.1, and $R_{\rm G}$ the BH gravitational radius
$R_{\rm G}=2G\mbh/c^2.$ Using the solution for the surface density in
the external region of the disk, $\Sigma(R)=\Sigma_0(R/R_{\rm
  G})^{-3/4}$ with

\begin{equation}\label{eqn:sigma}
\Sigma_0= 7\times 10^7\alpha_{0.1}^{-4/5}
M_{\rm BH,6}^{19/20} \left(\frac{f_{\rm Edd}}{\eta_{0.1}}\right)^{7/10}\,\rm {g\,cm^{-2}}, 
\end {equation}
the self-gravitating disk mass reads
\begin{equation}\label{eqn:msg1}
 m_{\rm sg}\approx (8\pi/5) R_{\rm G}^2\Sigma_0 (R_{\rm disk,sg}/R_{\rm G})^{5/4}, 
 \end{equation} and accordingly
\begin{equation}\label{eqn:msg2}
m_{\rm sg} \approx 2 \times 10^{4} \alpha_{0.1}^{-1/45}   \left(\frac{f_{\rm Edd}}{\eta_{0.1}}\right)^{4/45}M_{\rm BH,6}^{34/45}
\, \msun.  
\end{equation}

\begin{figure*}
\includegraphics[width=0.48\textwidth]{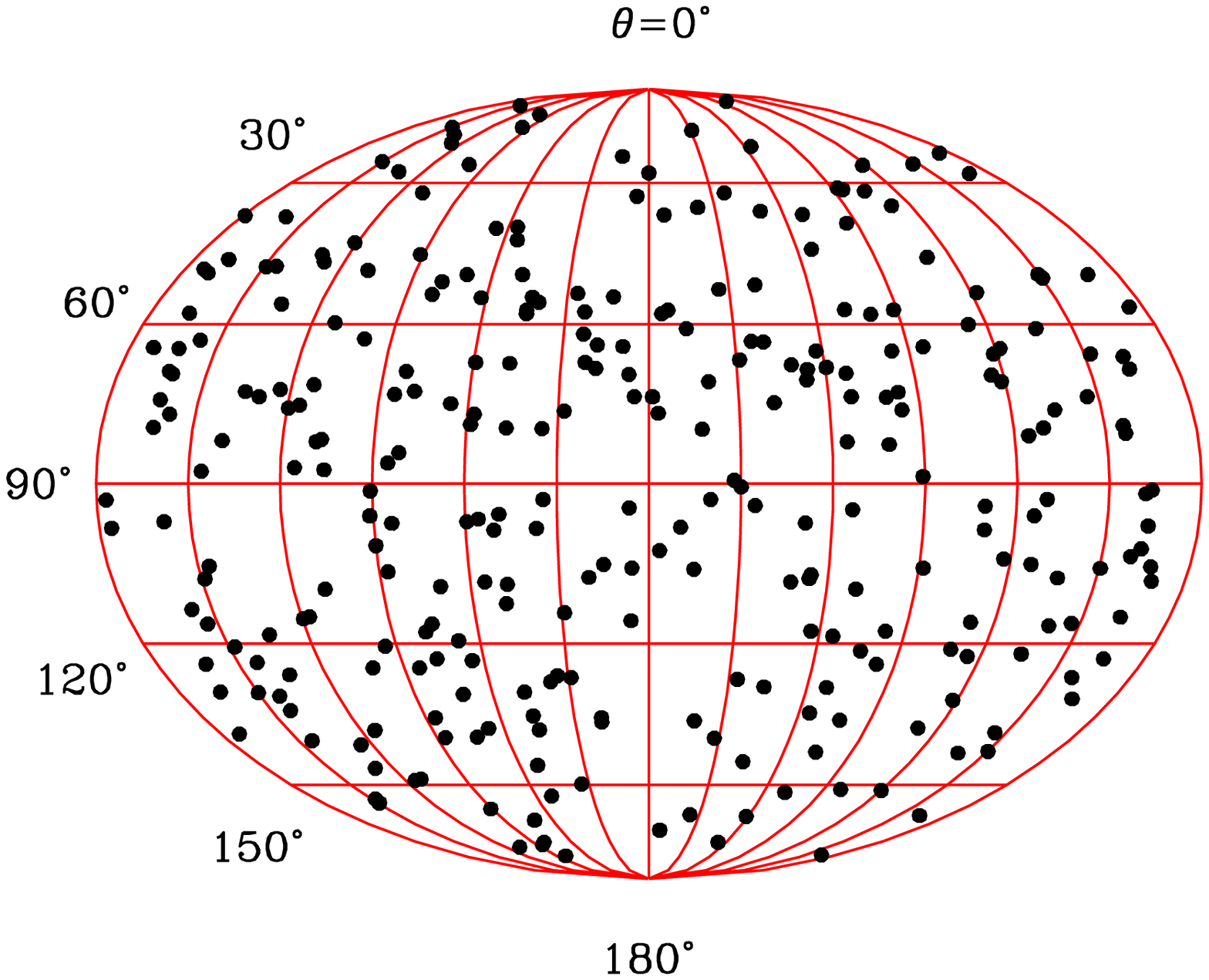}
\includegraphics[width=0.48\textwidth]{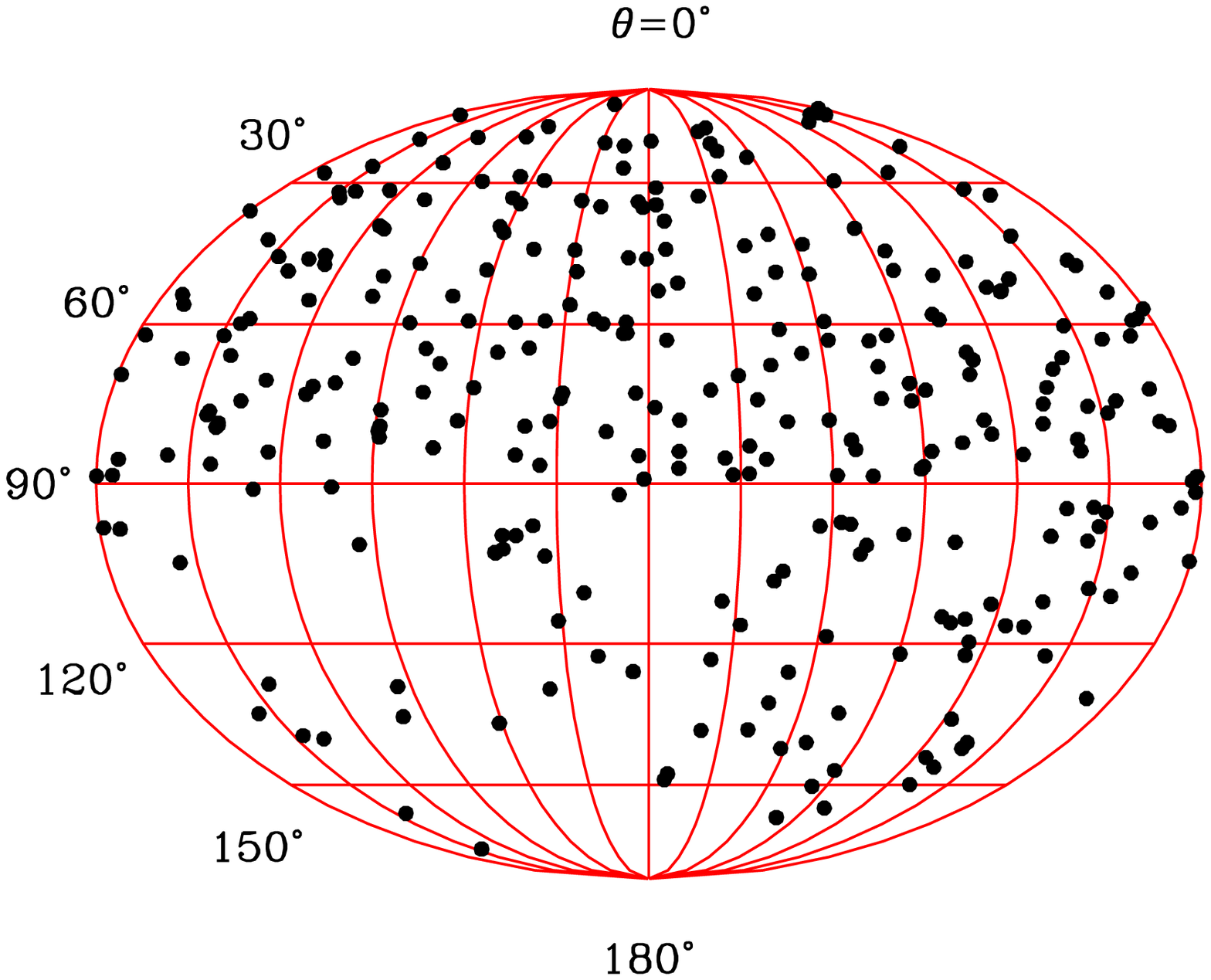}\\
\includegraphics[width=0.48\textwidth]{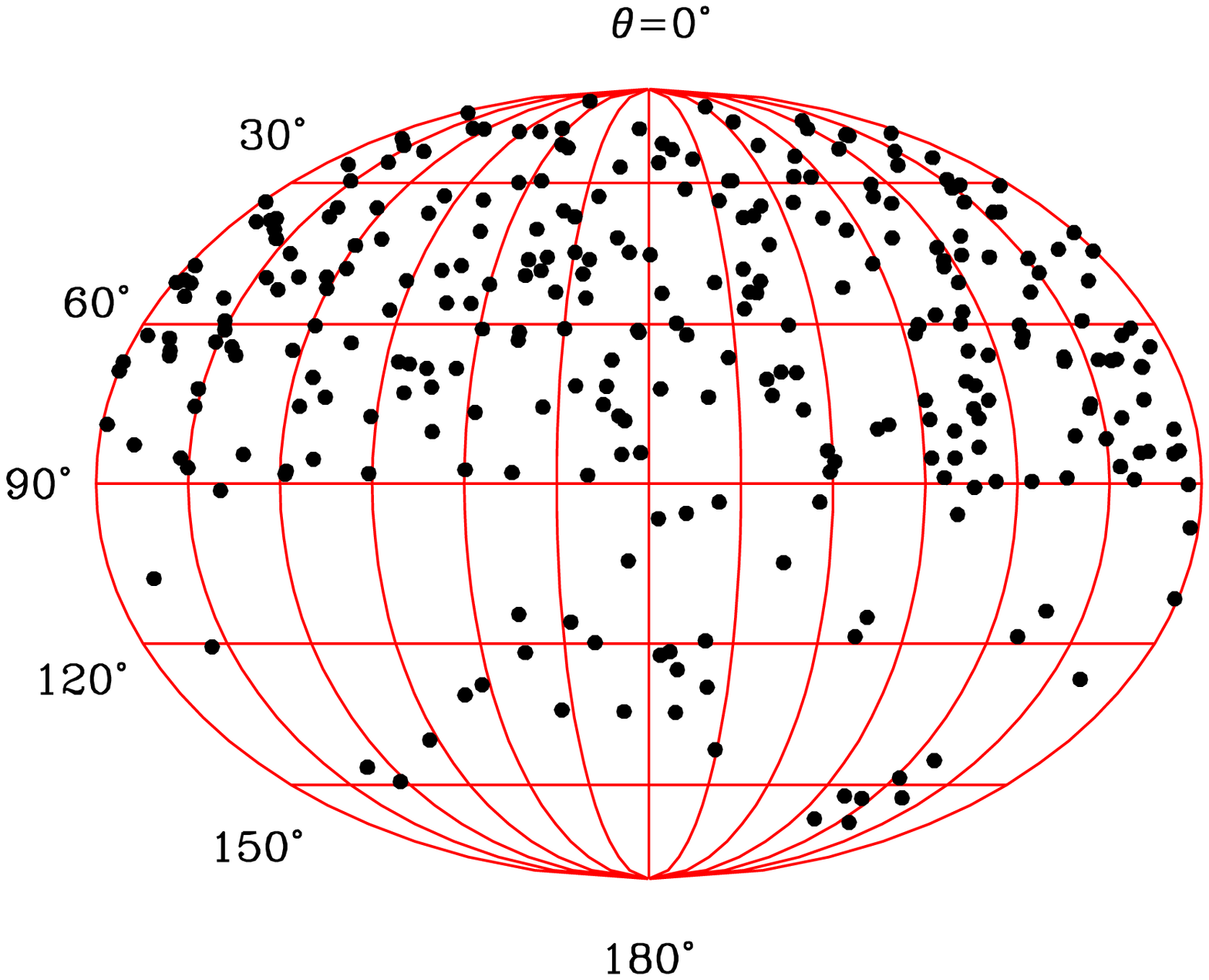}
\includegraphics[width=0.48\textwidth]{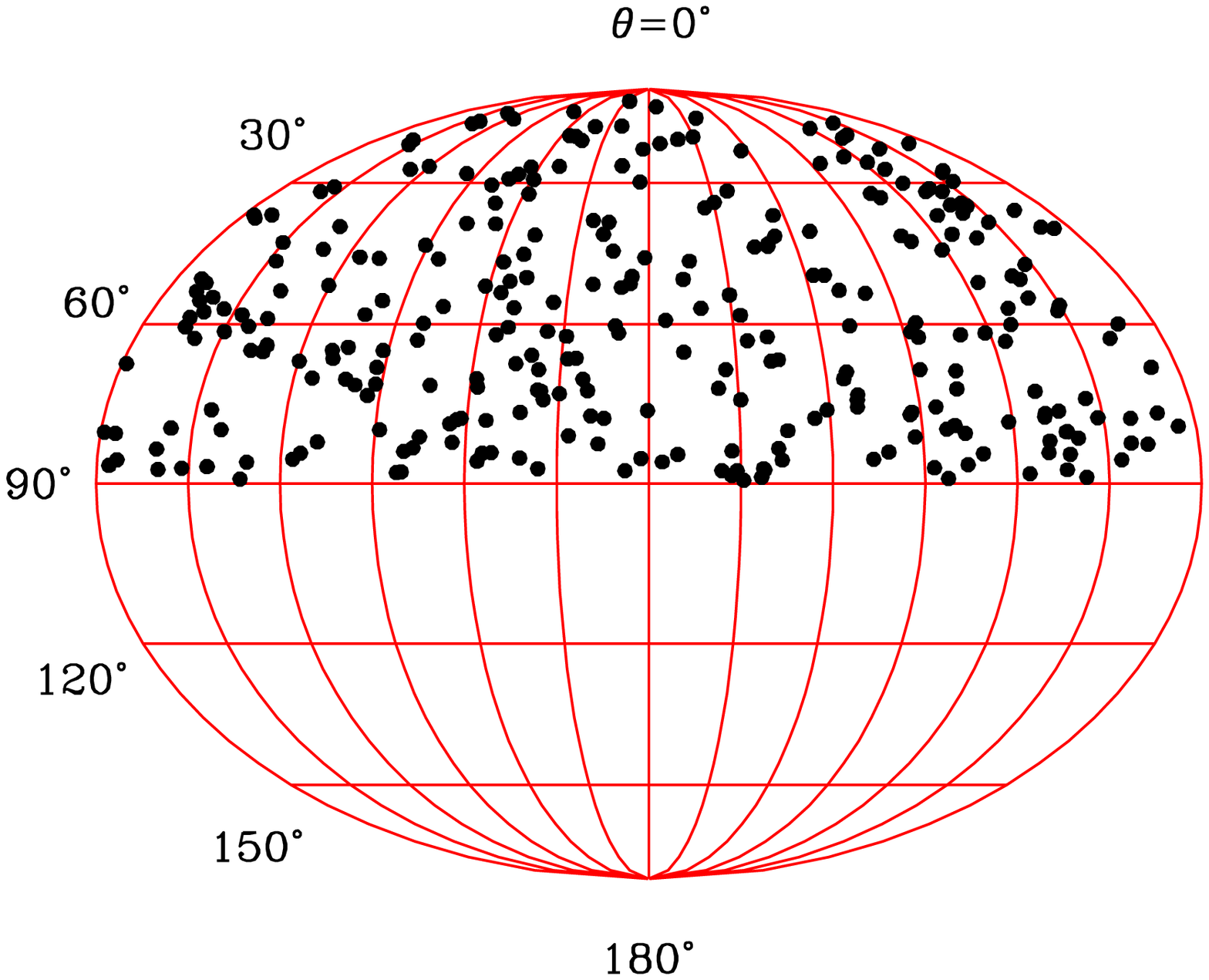}
\caption{Distribution of the directions of $\jdiskdir$ for different
  degrees of anisotropy. Top-left panel refers to the case of complete
  isotropy ("chaotic" accretion, $F=0.5$). Top-right, bottom-left, and
  bottom-right panels refer to accretion with levels of anisotropy
  $F=0.25, \, 0.125$ and 0, respectively.  }
\label{fig:maps}
\end{figure*}

When the disk mass equals $m_{\rm cloud}$, the outer radius $R_{\rm
  disk,cl}$ is computed as the integral over the surface density yielding,
\begin{equation}\label{eqn:rdiskcl}
{R_{\rm disk,cl}\over R_{\rm G}}\approx 4\times 10^4 \left ({m_{\rm cloud}\over 10^4\,\msun}\right )^{4/5}\alpha_{0.1}^{16/25}
M_{\rm BH,6}^{-44/25} 
\left (
{\fedd\over \eta_{0.1}}\right )^{-14/25}.
\end{equation}

Very massive BHs can stabilize huge accretion disks against their own
self-gravity. As an example, $m_{\rm sg} \lsim 10^7\,\msun$ for $\mbh
\sim 10^9\,\msun.$ Such a large reservoir of mass may not be available
in a single accretion episode.  In fact observations indicate
  that the mass fraction of cold gas relative to stars decreases with
  increasing galaxy mass (e.g. di Serego Alighieri et al. 2007,
  Catinella et al. 2010).  To avoid unrealistically massive accretion
episodes, we adopt the following criterion: if $m_{\rm cloud}>m_{\rm
  sg},$ the disk around the accreting BH is truncated at $R_{\rm
  disk,sg} $ and has mass $m_{\rm sg}$; by contrast if $m_{\rm
  cloud}<m_{\rm sg},$ the disk is truncated at $R_{\rm disk,cl}$ and
has mass $m_{\rm cloud}$. Thus for a given value of $m_{\rm cloud}$,
representing the mass reservoir in a single accretion episode, there
exists a characteristic BH mass $M^{\rm crit}_{\rm BH}$ for which
$m_{\rm sg}=m_{\rm cloud}:$
\begin{equation}\label{eqn:mcrit}
M^{\rm crit}_{\rm BH,6}=0.4 \alpha_{0.1}^{1/34}   \left(\frac{f_{\rm Edd}}{\eta_{0.1}}\right)^{-2/17}\left ({m_{\rm cloud}\over
10^4\,\msun}\right )^{45/34}.
\end{equation}

The mass during a single accretion episode is consumed over a
timescale $\tau_{\rm acc}$ that in the two regimes (when the BH mass
is below or above $M^{\rm crit}_{\rm BH}$ respectively) reads:

\begin {equation}\label{eqn:tauacc1}
\tau_{\rm acc,sg}={m_{\rm sg}\over {\dot M}}\approx 10^6 \alpha_{0.1}^{-1/45}M_{\rm BH,6}^{-11/45} \left(\frac{f_{\rm Edd}}{\eta_{0.1}}\right)^{-14/45}\, \rm yr,
\end{equation}

\begin {equation}\label{eqn:tauacc2}
\tau_{\rm acc,cl}={m_{\rm cloud}\over {\dot M}}\approx 5 \times 10^5 {m_{\rm cloud} \over 10^4 \msun} M_{\rm BH,6}^{-1} \left(\frac{f_{\rm Edd}}{\eta_{0.1}}\right)^{-1}\, \rm yr.
\end{equation}

\subsection{The black hole and the disk angular momenta}

The ratio of the angular momentum carried by the disk and the angular
momentum of the black hole is the main driver of how much an accretion
episode, and a series of accretion episodes can modify the direction
and magnitude of a BH spin.  The $\alpha-$disk carries an angular
momentum $\jdisk$ obtained integrating over all annuli the disk
angular momentum surface density $\bfl(R)=L(R)\ldir(R)$, where
$\ldir(R)$ is the local orientation of $\bfl$ in the disk, and
$L(R)=\Sigma(R) \Omega(R) R^2.$ Following Perego et al. (2009;
Perego09 hereafter),
\begin{equation}
\label{eqn:j disk R}
J_{\rm disk}(R)\propto \dot{M}\sqrt{GM_{\rm BH}}\,R^{7/4}.
\end{equation}
The ratio between the angular momentum of the disk and BH reads:
\begin{equation}\label{eqn:jdisk}
{{\rm J}_{\rm disk}(R_{\rm disk,sg})\over {\rm J}_{\rm BH}}\sim 7.3 \alpha_{0.1}^{13/45}\left (
{f_{\rm Edd}\over \eta_{0.1}}\right )^{-7/45}\, M_{\rm BH,6}^{-37/45}\, a^{-1}.
\end{equation}
Since ${\rm J}_{\rm disk}\propto
R^{7/4}$, disks that are not truncated by self-gravity
carry an angular
momentum smaller by a factor $(R_{\rm disk,cl}/R_{\rm disk,
  sg})^{7/4}$ than disks with mass $m_{\rm sg}$, and this will play a
role in modeling the BH history. 

The orientation of $\jdisk$, denoted as $\jdiskdir$, is determined
mainly by the angular momentum in the outermost region of the
disk. $\jdiskdir$ is uncorrelated with the BH spin direction
$\jbhdir$ at the onset of any accretion episode.  Hereon, disk
orientation is described in terms of the polar angle $\theta_{\rm
  disk}$ between $\jdiskdir$ and a fixed reference direction, i.e. the
unit vector ${\bf e}_z$ in our spherical coordinate reference system:
$\theta_{\rm disk}=\cos^{-1} (\jdiskdir \cdot {\bf e}_z)$.

The orientation of the accretion disk is selected through Montecarlo
sampling. We follow four different prescriptions that correspond to
different degrees of anisotropy in the BH fuelling process. In the
first, $\jdiskdir$ is distributed at random. This corresponds to a
uniform distribution for the azimuthal angle $\phi_{\rm disk}$ and to
a distribution proportional to ${\rm sin}\,\theta$ for the polar angle
$\theta_{\rm disk}$.  In this particular case, the fraction $F$ of
accretion events with $\theta_{\rm disk}>90^\circ$ is 0.5. The other
three cases are constructed following initially the same procedure,
but inverting at random the sign of $\theta_{\rm disk}$ to build three
distributions with $F=0.25$, $F=0.125$, and $F=0$. $F=0.25$
corresponds to a distribution with three times more events in the
``northern hemisphere'' ($\theta_{\rm disk}<90^{\circ}$) than in the
``southern hemisphere'' ($\theta_{\rm disk}>90^{\circ}$), $F=0.125$
corresponds to a distribution with a north-to-south events ratio of
seven, and the distribution with $F=0$ has no events in the southern
hemisphere. Figure~\ref{fig:maps} shows the distributions of
$\jdiskdir$ for the four values of $F$ considered.  Note that the
$F=0.5$ case is exactly what is assumed in the standard ``chaotic
accretion" scenario. $F=0$, on the other hand, does not correspond to
``coherent accretion'', but it mimics accretion through disk clouds
that are distributed isotropically but that share a common sense of
rotation.  $F=0$ corresponds to a 3D-dispersion to rotation velocity
ratio $\sigma/v_{\rm rot}\lsim 1$ for the gas fuelling the BH.

The direction of the BH spin vector $\jbhdir$, initially
selected at random, is followed by tracking the BH spin and accretion
history and by recording the values of $a$, $\jbhdir$ and $\mbh$ at
the end of each accretion episode.  Hereon we denote with $\thetabh$
the polar angle of $\jbhdir$ with ${\bf e}_z$,
i.e. $\thetabh=\cos^{-1}(\jbhdir\cdot{\rm e}_{z})$.  The BH spin
vector can point in all directions, and $\jbhdir$ as well as
$\jdiskdir$ are referred to the reference frame of the galaxy defined
by ${\bf e}_{z}$. 

\subsection{$a$, $\jbhdir$ and $\mbh$ on a history path}\label{code}

At the onset of any accretion episode the BH spin vector $\jbh$ is
generally misaligned with respect to the direction of the accretion
disk $\jdiskdir$.  In high viscosity $\alpha-$disks, this
configuration is unstable and evolving into a lower energy state.
There is an early phase in which gravito-magnetic torques exerted by
the spinning BH on disk's fluid elements cause the disk to warp on a
timescale $\tau_{\rm warp}$ (Bardeen \& Petterson 1975); the second
phase is the alignment phase, i.e. a change in the orientation of the
BH spin over a longer time.

In the early phase, the spinning BH induces Lense-Thirring precession
of the orbital plane of disk's fluid elements: precession of the plane
occurs at a frequency $\omega_{\rm prec}=(2G/c^2) {\rm J}_{\rm
  BH}/R^3=(2G/c) a \mbh^2/R^3$, so that fluid elements closer to the
BH precess faster.  Close to the BH, their orbital plane tends to
align (or anti-align if counter-rotating) parallel to the BH spin
direction $\jbhdir,$ on times shorter at shorter radii, so that the
perturbation diffuses radially outwards, on a timescale $\tau_{\rm
  warp}\sim R^2/\nu_2$, where $\nu_2$ is the vertical shear viscosity.
The disk is maximally warped at  $R_{\rm warp},$ corresponding
to the distance where the warp-diffusion timescale becomes comparable
or shorter than the precession time $\omega_{\rm prec}^{-1}$:
\begin{equation}\label{eqn:rwarp}
{R_{\rm warp}\over R_{\rm G}}\sim {4G{\rm J}_{\rm BH}\over \nu_2c^2R_{\rm G}}\sim 500 \alpha_{0.1}^{24/35}
f_{\nu_2}^{4/7}M_{\rm BH,6}^{4/35} \left(\frac{f_{\rm Edd}}{\eta_{0.1}}\right)^{-{6\over 35}}a^{4/7}, 
\end{equation}
where $\nu_2\sim \alpha_2 c_{\rm s}H$ with $\alpha_2\approx
f_{\nu_2}/(2\alpha)$ (Lodato \& Pringle 2007; Perego09 for details)
$R_{\rm warp}$ indicates
the region in the inner disk where the fluid elements align or
antialign with $\jbhdir$.  At $R\sim R_{\rm warp}$ the warp timescale
is

\begin{equation}\label{eqn:tauwarp}
\tau_{\rm warp}\sim 35\, \alpha_{0.1}^{72/35} f_{\nu_2}^{-12/7}M_{\rm BH,6}^{47/35}\left(\frac{f_{\rm Edd}}{\eta_{0.1}}\right)^{-{18\over 35}}
a^{5/7}\,\,\rm yr.
\end{equation}
Since the warp timescale  is shorter than the viscous timescale at all annuli, the
deformation diffuses more rapidly outwards than the inward radial drift motion so that
the deformed disk attains an equilibrium profile, and the shape of the
perturbation is stationary (Martin et al. 2007).

The warped disk (in the small deformation approximation) is described
by an equilibrium surface density $\Sigma(R)$ and a velocity
$\Omega(R)$ close to that of an unperturbed Keplerian disk, plus a
deformation in the local angular momentum vector (per unit surface
area) ${\mathbf L}=L(R)({\hat l}_x,{\hat l}_y,{\hat l}_z)$ (with
$\bflhat$ a unit vector in the direction of $\bfl$; we defer to
Perego09 for details).
Angular momentum conservation then
imposes the spinning BH to precess and align relative to the total
angular momentum $\jtotal=\jdisk+\jbh$.  During individual accretion
episodes at a rate $\dot M$, the BH mass increases according to:

\begin{equation}\label{eqn:massevol}
{d \mbh\over dt}=(1-\eta){\dot M}c^2=(1-\eta)\left(\frac{f_{\rm Edd}}{\eta}\right){\mbh\over {\tau_{\rm S}}},
\end{equation}
with an $e-$folding timescale
\begin{equation}\label{eqn:taumass}
\tau_{\mbh}\sim {\eta {\tau_{\rm S}}\over (1-\eta)\fedd}\sim 5\times 10^7 {\eta_{0.1} \over (1-\eta_{0.1})\fedd}\,\rm yr.
\end {equation}
The BH angular momentum vector evolves according to
\begin{equation} \label{eqn:jbh}
\frac{d{\mathbf J}_{\rm BH}}{dt} =\dot{M}{G\mbh\over c}\Lambda{_{\rm isco}}\hat
 {\mathbf l}(R_{\rm ISCO}) + \frac{4\pi G}{c^2}\int_{\rm
 disk}\frac{{\mathbf L} \times {\mathbf J}_{\rm BH}}{R^2}dR
 \end{equation}
where the first term in equation (15) describes the change in the spin
modulus $a$ due to the transfer of angular momentum per unit mass at
the innermost stable circular orbit $R_{\rm ISCO}$ (where
$\Lambda(R_{\rm ISCO})$ is a known dimensionless function of
$a$). Gravito-magnetic coupling in the inner region of the precessing
disk ensures that the direction of $ {\bflhat}(R_{\rm ISCO})$ is
parallel o anti-parallel to $\jbhdir$.  
 
The second term of equation~(\ref{eqn:jbh}) describes the change in the BH
spin orientation which tends to reduce the degree of misalignment between the
disk and the BH spin (Perego09; Scheuer \& Feiler 1996).  An intuitive
explanation for the alignment is as follows.  The warping of the disk results
always in a either
aligned or antialigned inner part of the accretion disk,
due to the BH metric.  On the other hand, since the frequency of the
Lense-Thirring precession decreases with
radius, the outer part of the
accretion disk can effectively be
considered unperturbed (this corresponds to
the Newtonian limit for
the accretion disk), and there exists a region (near
the so called
warp radius) where the local disk angular momentum unit
vector
is strongly misaligned both with the spin and disk's angular
momentum
at far distances (in the unperturbed disk). In other words, gas
moving from 
the unperturbed outer region into the innermost region modifies
its
angular momentum direction. As a consequence, the BH has to (at least
partially)
align with the direction of the angular momentum of the
outer
accretion disk to ensure conservation of the total angular momentum,
and
to compensate the "loss/change" of angular momentum in the warp region.
The alignment effect on the BH spin due to the cumulative
torques exerted by
a mass distribution does not necessarily require
the presence of
viscosity. The same alignment can be induced (under
some conditions) by the
spin orbit interaction between the BH and a 
rotating stellar cusp (see
Merritt \& Vasiliev 2012).
Viscosity guarantees alignment of the inner disk
region and thus 
guarantees the occurrence of a stationary warp region in the
disk causing
the torquing of the BH and in turn spin alignment.  
  From
  equation~(\ref{eqn:jbh}), we can infer an evolution equation for the spin
  modulus
  \begin{equation}\label{eqn:a}
 {d a\over dt}=\left [ {\Lambda(R_{\rm ISCO})\over \eta}-2a\left({1\over
     \eta}-1\right)\right ]{ f_{\rm Edd}\over \tau_{\rm S}}.
 \end{equation}
Equation~(\ref{eqn:a}) shows that changes in the spin modulus $a$ typically
occur on the timescale $\tau_{\rm spin}=a/{\dot a}$ comparable to
$\tau_{\mbh}$ (eq.~\ref{eqn:taumass}). As shown in Bardeen (1970), a
non-spinning BH ($a=0$) is spun up to an extreme Kerr BH ($a=1$) after having
accreted a mass $\sqrt{6}\mbh.$ The timescale of the alignment process
(\ref{eqn:jbh}) is much shorter.  The peak of the torque perpendicular to
the
BH spin (that responsible for the evolution of the spin direction)
is maximal
at the warp radius (where the direction of the angular momentum
of the gas in
the disk changes). The warp radius is considerably
larger than the last
stable orbit, so the angular momentum per unit of
mass of the gas at the warp
radius is much larger than that at the
ISCO. As a consequence, the torque
exerted onto the BH responsible for
the evolution of its direction is
considerably larger than that
responsible for the spin magnitude
evolution. Consequently, this results in
a shorter timescale for the spin
alignment with respect to that
associated to the spin magnitude evolution.

Equations~(\ref{eqn:massevol}) and (\ref{eqn:jbh}) are integrated
considering initial BH masses of $M_{\rm BH,0}=10^5\,\msun$, and
arbitrary initial spin moduli ($a_0$) and orientations; $\fedd$ that
enters all timescales can be considered as a scaling parameter so that
smaller values of $\fedd$ imply longer times. In the calculation we
fix $\fedd=0.1$.  According to equation~(\ref{eqn:jbh}), if the disk
angular momentum contributes most to $\jtotal$, the BH spin-disk
orientation angle $\zeta_{\rm BH,disk}=\cos^{-1}(\jbhdir\cdot
\jdiskdir)$ reduces to zero (i.e. full alignment even starting with
$\zeta_{\rm BH,disk}=180^{\circ}$) on a timescale
 \begin{equation}\label{eqn:taualign}
 \tau_{\rm al} \sim 10^5 \alpha_{0.1}^{58/35}f_{\nu_2}^{-5/7} M_{\rm BH, 6}^{-2/35}
 \left ({ f_{\rm Edd}\over \eta_{0.1}}\right )^{-{32\over35}}
 a^{5/7} {\rm
   yr}  
\end{equation} 
(Martin et al. 2007; Lodato \& Pringle 2006; Perego09). 
The timescales $\tau_{\rm warp}$ and $\tau_{\rm al}$ depend on $a$ and
$\mbh$ explicitly, and the following inequalities hold
\begin{equation}\label{eqn:tauchain}
\tau_{\rm warp}<\tau_{\rm al}<\tau_{\rm acc}<\tau_{\mbh}\sim \tau_{\rm spin}.
\end{equation}
In our study, we integrate the BH spin evolution equation over
timesteps $\delta t$ that are $\tau_{\rm warp}<\delta t<\tau_{\rm al}$
tracing contemporarily the change in $a$, $\jbhdir$ and $\mbh$.  We
notice that as $\tau_{\rm spin} > \tau_{\rm al}$, the spin
orientation changes rapidly with time, so that we can exclude the
possibility that retrograde accretion reduces $a$ to 0 before
re-increasing it to 1.  The rapid change in $\jbhdir$ compared to the
change in $a$, favors conditions of prograde accretion episodes, even
starting from retrograde conditions.

The progressive increase in $\mbh$ during accretion episodes with
constant $m_{\rm disk}=m_{\rm cloud}$ (when $\mbh>\mbh^{\rm crit}$)
implies the formation of an accretion disk with $m_{\rm disk}/\mbh$
always decreasing.  Thus the disks that form during single accretion
episodes are progressively smaller and carry less angular momentum
relative to the BH. Under these circumstances and depending on $a$,
the dimensionless ratio ${\rm J}_{\rm disk}/{\rm J}_{\rm BH}\lsim 1$,
and the gravito-magnetic BH-disk coupling has little influence on the 
BH spin direction.   At even larger masses
(already in the regime where ${\rm J}_{\rm disk}/{\rm J}_{\rm BH}< 1$)
the warp radius $R_{\rm warp}$ may rise above $\gsim R_{\rm disk}$
depending on $a$, and equation~(\ref{eqn:jbh}) becomes invalid (Martin et al. 2007).  
This transition occurs when $M_{\rm BH,6}> M^{\rm warp}_{\rm BH,6}$, where 
\begin{equation}\label{eqn:warpmass}
M^{\rm warp}_{\rm BH,6}\approx10 \,\left ({m_{\rm cloud}\over 10^4\,\msun}\right )^{35/82}\alpha_{0.1}^{-1/41}a^{-25/82}\left(\frac{f_{\rm Edd}}{\eta_{0.1}}\right)^{-{17\over 82}}.
\end{equation}
Under these circumstances, we assume that in any single event $\jbh$
aligns instantaneously to the direction of the total angular momentum
$\jtotal$ (King et al. 2005). The re-orientation of $\jbh$ is rather
small, since in this regime the BH spin is quite close to the
  total angular momentum already at the beginning of every individual
  accretion episode ($\jtotal$ is in fact dominated by $\jbh$). $\jdisk$ instead undergoes a fast and significant
  re-orientation, and is either aligned or antialigned with respect to
  $\jbh$ depending on the initial angle of relative misalignment
  ($\zeta_{\rm BH,disk}$) and the ${\rm J}_{\rm disk}/{\rm J}_{\rm
    BH}$ ratio.  The two angular momenta are aligned if
$\cos(\zeta_{\rm BH,disk})>-{\rm J}_{\rm disk}/2 {\rm J}_{\rm BH}$ (or
counter-aligned when the opposite inequality holds; King et al. 2005;
Lodato \& Pringle 2006).  The BH mass and spin parameter are then
evolved according to equation~(\ref{eqn:massevol}) and (\ref{eqn:a}).

To summarize, our procedure leads to the following
  sequence:\\ $\bullet$ The disk mass is the minimum between the cloud
  mass and the self-gravitating mass, $m_{\rm disk}=\min(m_{\rm
    cloud},m_{\rm sg})$. The transition between the two phases occurs
  when the BH mass reaches the value $M_{\rm BH,6}^{\rm crit}$ given by equation~(\ref{eqn:mcrit}); 
   then the disk contains the whole cloud mass.

\noindent
$\bullet$ Up until $M_{\rm BH,6}<M^{\rm warp}_{\rm BH,6}$, we follow the
 spin-disk alignment as a function of time by solving for equation~(\ref{eqn:jbh}).
 Here, $M^{\rm warp}_{\rm BH,6}$ is given by equation~(\ref{eqn:warpmass}).
 Above this mass the disk is assumed to be aligned or anti-aligned with
 the angular momentum of the BH, depending on the initial angle of relative misalignment.

Therefore, three regimes exist, $M_{\rm BH,6}<M^{\rm crit}_{\rm
  BH,6}$; $M^{\rm crit}_{\rm BH,6}<M_{\rm BH,6}<M^{\rm warp}_{\rm
  BH,6}$, and $M_{\rm BH,6}>M^{\rm warp}_{\rm BH,6}$. While the switch
between the first and second regime depends only on the properties of
the gas fuelling the BH, the switch between the second and third
regime depends also on the BH spin, and this threshold mass,
therefore, depends on the whole growth history of the BH, i.e., how it
gained its spin. We anticipate that, regardless of the regime, the
important parameter that determines BH spin evolution is the ratio
${\rm J}_{\rm disk}/{\rm J}_{\rm BH}$, as previously discussed.

 \section{Results}

\begin{figure*}
\includegraphics[width=0.96\textwidth]{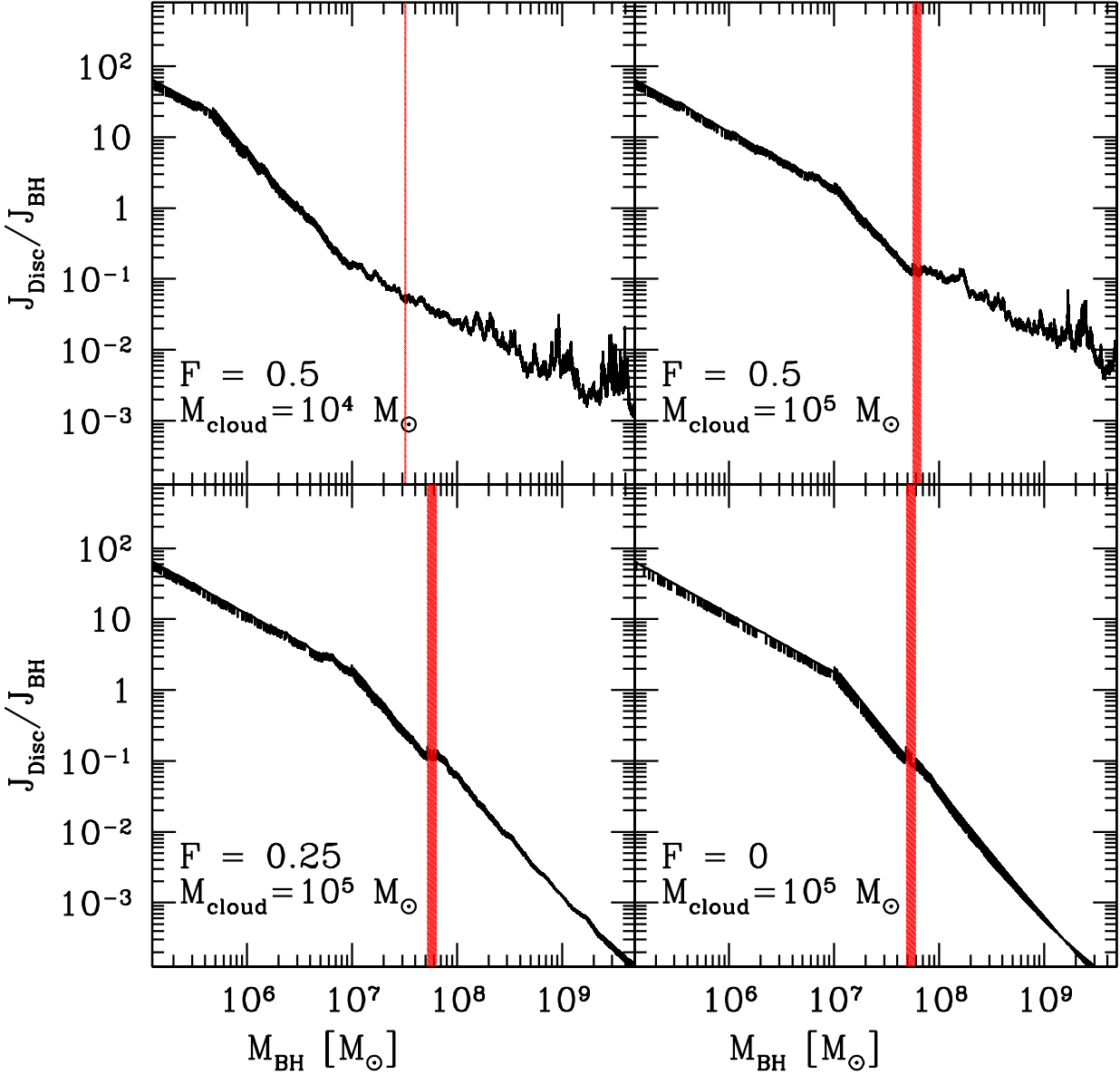}
\caption{Ratio between the total angular momentum of the accretion
  disk $\jdisk$ and the BH spin $\jbh$, as a function of the BH mass
  $\mbh$, along an accretion history. The upper left panel refers to
  the isotropic case ($F=0.5$) with a maximum mass per accretion event
  $m_{\rm cloud}= 10^4\msun$. The results in the upper right, lower
  left and lower right panels assume $m_{\rm cloud}=10^5 \msun$ and
  $F=0.5$, 0.25, and 0, respectively. Three regimes exist, $M_{\rm
    BH,6}<M^{\rm crit}_{\rm BH,6}$; $M^{\rm crit}_{\rm BH,6}<M_{\rm
    BH,6}<M^{\rm warp}_{\rm BH,6}$, and $M_{\rm BH,6}>M^{\rm
    warp}_{\rm BH,6}$. The knee at low BH masses shows the transition
  between the first and the second regime.  The second transition
    is highlighted by the shaded vertical area (see text for details).}
\label{fig:jdisk_over_jbh}
\end{figure*}

Figure~\ref{fig:jdisk_over_jbh} shows the evolution of the ratio
between ${\rm J}_{\rm disk}$ and ${\rm J}_{\rm BH}$ as a function of
the BH mass, for accretion histories that differ in the value of
$m_{\rm cloud}$ and of the anisotropy parameter $F$.

As discussed in the previous section we can distinguish an early phase
(I) during which the accretion disks carry a large angular momentum,
and a second phase (II)
where the opposite holds.  During phase I, the disks are initially truncated by
their own self-gravity and their mass is determined
by the BH mass.  As the BH grows in mass the disks carry the cloud mass 
$m_{\rm cloud}$.  This transition
occurs at $M_{\rm BH}^{\rm crit}$ and is visible as a knee in the
${\rm J}_{\rm disk}/{\rm J}_{\rm BH}$ versus BH mass diagram.  During phase
II, disks are tiny ($R_{\rm warp}>R_{\rm disk}$) and there is the switch between the two prescriptions for the BH spin evolution.
This occurs around a few $10^7\,\msun$ and is highlighted as a red shaded
area in each panel.  The upper left and right panels refer to the
isotropic case ($F=0.5$) with a maximum mass per accretion event
$m_{\rm cloud} = 10^4\,\msun$ and $10^5\,\msun$, respectively.  The
minimum value of ${\rm J}_{\rm disk}/{\rm J}_{\rm BH}$ for the two
cases is between $10^{-2}-10^{-3}$ and is attained when $\mbh\gsim
10^9\,\msun.$ Similar results are found for different values of the
anisotropy parameter, as shown in the bottom panels for $m_{\rm cloud}
= 10^5\,\msun$, assuming $F=0.25$ and $0$ respectively.  Notice that
in these cases ${\rm J}_{\rm disk} /{\rm J}_{\rm BH}$ drops below
$10^{-3}$ for very large masses and this reflects the large spin
carried by these BHs.  For anisotropic cases ($F=0.25; 0$) the massive
BHs tend to align their spins with the average angular momentum of the
gas reservoir, as will be discuss in
section~\ref{sec:spindirection}. This results in an higher fraction of
prograde accretion events, and, as a consequence, in a higher ${\rm
  J}_{\rm BH}$. A detailed description of the evolution in direction
and magnitude of $a$ is presented in the next two sections.

\subsection{Evolution of the BH spin direction}\label{sec:spindirection}

\begin{figure*}
\includegraphics[width=0.96\textwidth]{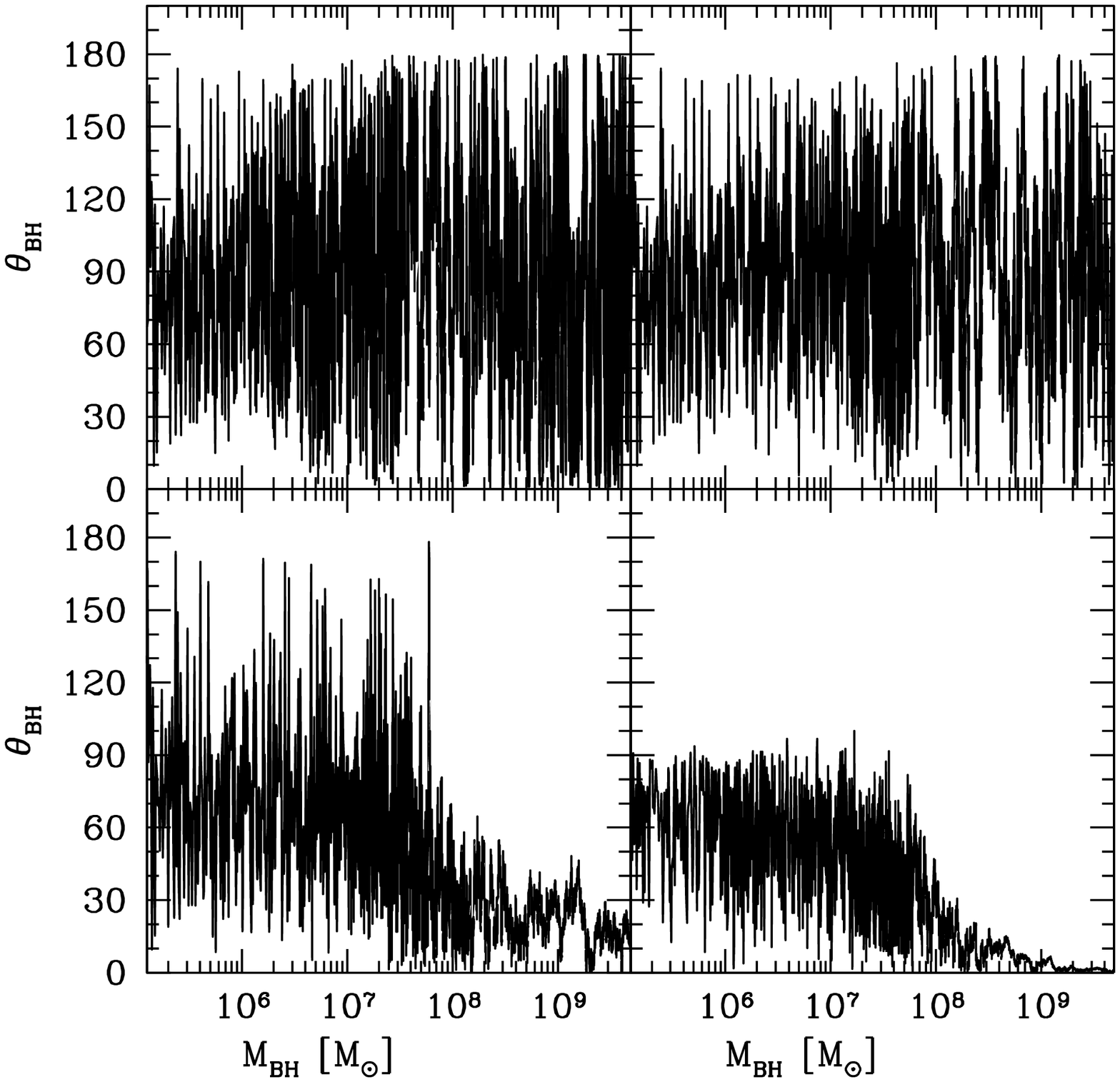}
\caption{Evolution of the polar angle $\theta_{\rm BH}$ describing the
  BH spin orientation, relative to a fixed coordinate system, as a
  function of the BH mass.  The upper left panel refers to the
  isotropic case ($F=0.5$) with a maximum mass per accretion event
  $m_{\rm cloud} = 10^4\msun$. The results in the upper right, lower
  left and lower right panels assume $m_{\rm cloud}=10^5 \msun$ and
  $F=0.5$, 0.25, and 0, respectively.}
\label{fig:theta_single}
\end{figure*}

\begin{figure*}
\includegraphics[width=0.96\textwidth]{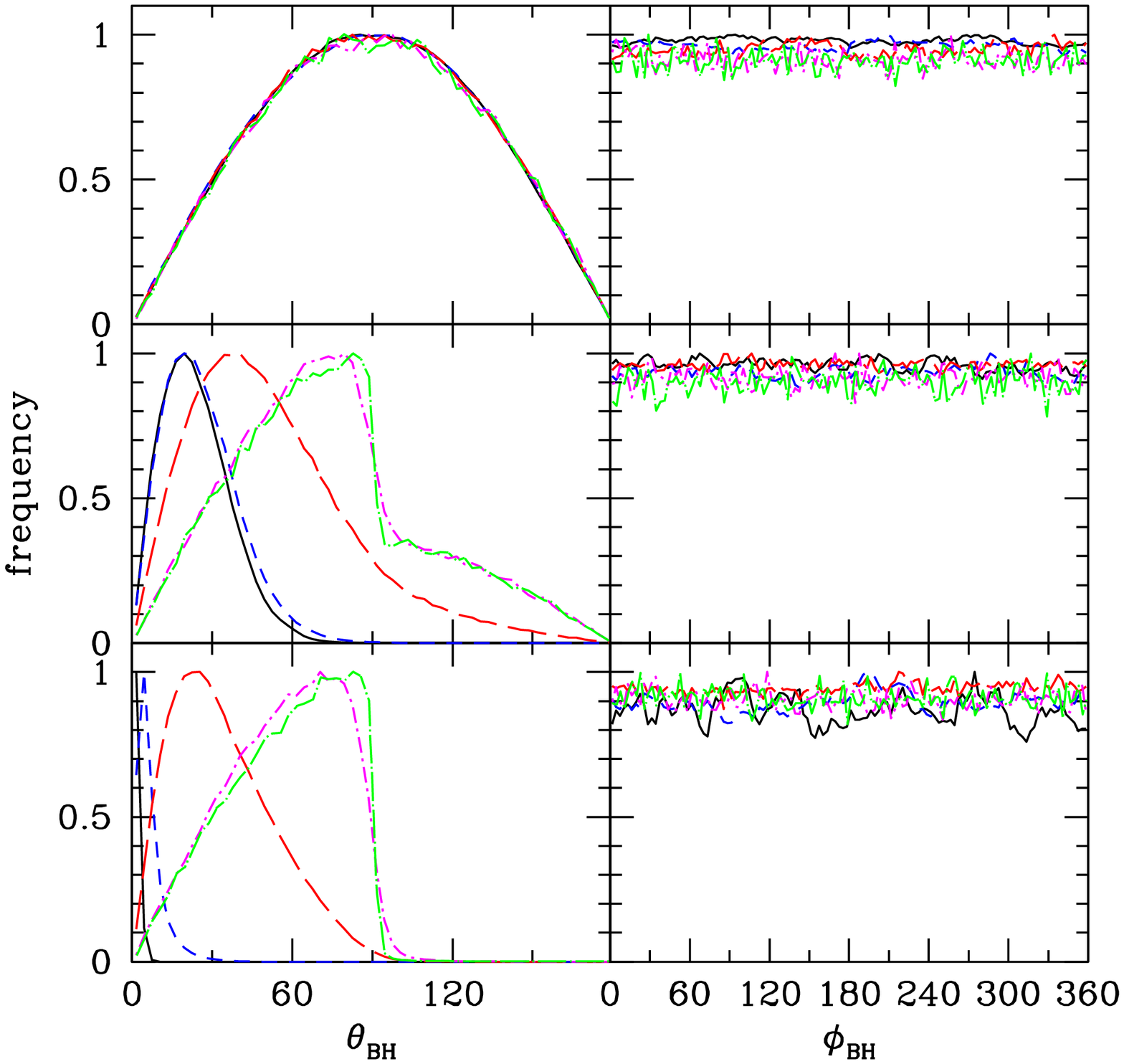}
\caption{Distributions of the polar angle $\theta_{\rm BH}$ (left
  column) and azimuth angle $\phi_{\rm BH}$ (right column), for BHs of
  different masses, assuming $m_{\rm cloud}=10^5\, \msun$. Top,
  middle, and bottom panels refer to the $F=0.5$, 0.25, and 0 cases,
  respectively.  BHs in the mass interval $10^5\,\msun -10^6\, \msun$
  are denoted with green, long dot-dashed line, $10^6\,\msun - 10^7\,
  \msun$ with purple, short dot-dashed line, $10^7\,\msun - 10^8
  \,\msun$ with red, long-dashed line, $10^8 \,\msun- 10^9\, \msun$
  with blue, short-dashed line, and $ > 10^9\, \msun$ with black,
  solid line.  }
\label{fig:theta_distribution}
\end{figure*}

The BH spin orientation in a fixed reference frame is defined by two
angles, the polar angle $\theta_{\rm BH}$, and the azimuthal angle
$\phi_{\rm BH}$.  Figure~\ref{fig:theta_single} shows, as an example,
the evolution of $\theta_{\rm BH}$ as a function of the BH mass for
different accretion histories with $F=0.5$ (upper panels) $0.25$ and
0 (lower left and right panels respectively), and cloud masses $m_{\rm
  cloud}=10^4 \msun$ (upper left panel), and $10^5 \msun$ (other
panels).  In case of perfect isotropy in the BH fueling ($F=0.5$;
upper panels in Figure~\ref{fig:jdisk_over_jbh}), the lack of any
preferential direction results in an unbiased random walk of the BH
spin direction. This can be quantitatively viewed by plotting the
distributions of $\theta_{\rm BH}$ and $\phi_{\rm BH}$ averaged over
500 accretion histories extracted at random. In the upper panels of
Figure~\ref{fig:theta_distribution} we show the distribution of the
two angles for the isotropic case for BHs in the mass interval
$10^5\,\msun - 10^6 \, \msun$ (green, long dot-dashed line), $10^6
\,\msun - 10^7\, \msun$ (purple, short dot-dashed line), $10^7\,\msun
- 10^8\, \msun$ (red, long dashed line), $10^8\,\msun - 10^9\, \msun$
(blue, short dashed line) and $ > 10^9 \msun$ (black, solid line).  It
is clear that both distributions are consistent with isotropy.

The lower-left panel of Figure~\ref{fig:theta_single} refers to the
case when $F=0.25$, corresponding to 3/4 of the accretion episodes
having the disk angular momentum confined in the northern hemisphere.
The behavior of $\theta_{\rm BH}$ differs in the two accretion phases
(described in the previous section).  When ${\rm J}_{\rm disk}/{\rm
  J}_{\rm BH}>1$ (for masses $\mbh\lsim {\rm few}\,\, 10^7\,\msun$)
the spin orientation aligns significantly with the angular momentum of
{\it each accretion episode}, thus changing erratically over the full
range ($0,180^{\circ}$, despite the fact that the mean is less than
$90^{\circ}$). By contrast, during phase II (${\rm J}_{\rm disk}/{\rm
  J}_{\rm BH}<1$) the spin tends to align with the {\it average
  angular momentum} of the accreting material. In this phase the
  BH spin slightly tilts toward the direction of the angular momentum
  of each accreting cloud. The sequence of accretion episodes results
  in a random walk biased toward the direction of the average angular
  momentum of the inflow. This trend is illustrated in the two
middle panels of Figure~\ref{fig:theta_distribution}.  While
$\phi_{\rm BH}$ retains a flat distribution, the distribution of
$\theta_{\rm BH}$ covers the full $[0,180^{\circ}]$ range for small BH
masses, with a 3 times higher probability of $\theta_{\rm
  BH}<90^{\circ}$. For higher masses (i.e. during phase II), the
distribution of $\theta_{\rm BH}$ shifts toward lower angles, being
confined below $\sim 60^{\circ}$ for $\mbh \gsim 10^8 \msun$.
Similarly, for the most anisotropic case considered here ($F=0$), the
BH spin is confined within a solid angle that decreases with
increasing BH mass (see the lower right panel of
Figure~\ref{fig:theta_single} and the lower two panels of
Figure~\ref{fig:theta_distribution}).  For the very massive BHs
($\mbh>10^9\,\msun$), $\jbhdir$ is nearly constant, demonstrating the
stability of the spin direction even during substantial increase in
mass (from $10^9$ to $5 \times 10^9\,\msun$).

In summary, we show that for small BH masses ($\mbh\lsim 10^7\,\msun$)
the BH spin always aligns to the disk angular momentum in every single
accretion episode, and can be substantially misaligned relatively to
the average of the angular momenta of the disks.  For $\mbh\gsim
10^7\,\msun$, a single accretion episode does not modify significantly
the BH spin direction. In this regime, the BH spin aligns with the
average direction of the angular momentum of the accreting material,
with a degree of alignment that increases with increasing anisotropy
in the fueling process.

\subsection{Evolution of the BH spin magnitude $a$}\label{a}

In this section we focus on the evolution of the spin magnitude $a$
considering for the first time its coupling with the orientation
$\jbhdir$ and with the dynamical properties of the accretion.

\begin{figure}
\includegraphics[width=0.48\textwidth]{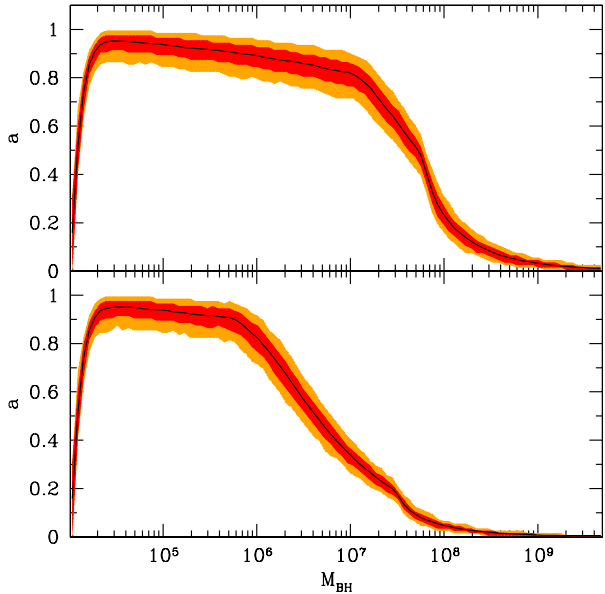}
\caption{Evolution of the BH spin magnitude $a$ as a function of
  $M_{\rm BH}$, for the isotropic case ($F=0.5$). The upper (lower)
  panel refers to accretion episodes with $m_{\rm cloud} = 10^5 \msun$
  ($10^4\msun$).  The black line refers to the mean over 500
  realizations. Red and orange shaded areas enclose intervals at
  $1-\sigma$ and $2-\sigma$ deviations, respectively.  }
\label{fig:a_cloud}
\end{figure}

Figure~\ref{fig:a_cloud} shows the evolution of the BH spin $a$ as a
function of its mass, for the isotropic case and two different values
of $m_{\rm cloud}$.  The memory of the initial spin is erased after
the BH accretes a few times its initial mass ($M_{\rm
  BH,0}=10^4\,\msun$), and $a$ rises attaining values $\approx 0.9$
during phase I.  The large spin is a consequence of the rapid
alignment induced by the Bardeen-Petterson effect that occurs on a
timescale $\tau_{\rm al} < \tau_{\rm acc} < \tau_{\rm spin}$ turning
initially retrograde accretion into prograde accretion before disk
consumption.  Later, i.e. for larger BH masses, the spin drops to
lower values, down to $a\approx 0,$ for $\mbh\gsim 10^9\,\msun$. The
beginning of the spin decrease corresponds to the transition between
phase I and II that occurs at different BH masses, depending on the
value of $m_{\rm cloud}$, as shown in the figure.  The parameter that
controls this transition is ${\rm J}_{\rm disk}/{\rm J}_{\rm BH}$:
smaller clouds carry lower angular momenta and the transition between
phase I and II occurs at smaller BH masses.  The spin modulus
decreases because over a single accretion episode $\jbhdir$ does not
change significantly. Retrograde accretion episodes remain retrograde
over the disk consumption timescale, so that, for isotropic
  fueling, the probability of having  a prograde accretion events is
  exactly the same of having a retrograde one.   Because the location 
  of the ISCO is farther away, and therefore the accreted material carries a 
  larger specific angular momentum,  retrograde
accretion transfers more angular momentum per unit mass than prograde
accretion, isotropic fueling results in net spin-down and thus low
spins (e.g. King \& Pringle 2006).

\begin{figure*}
\includegraphics[width=0.96\textwidth]{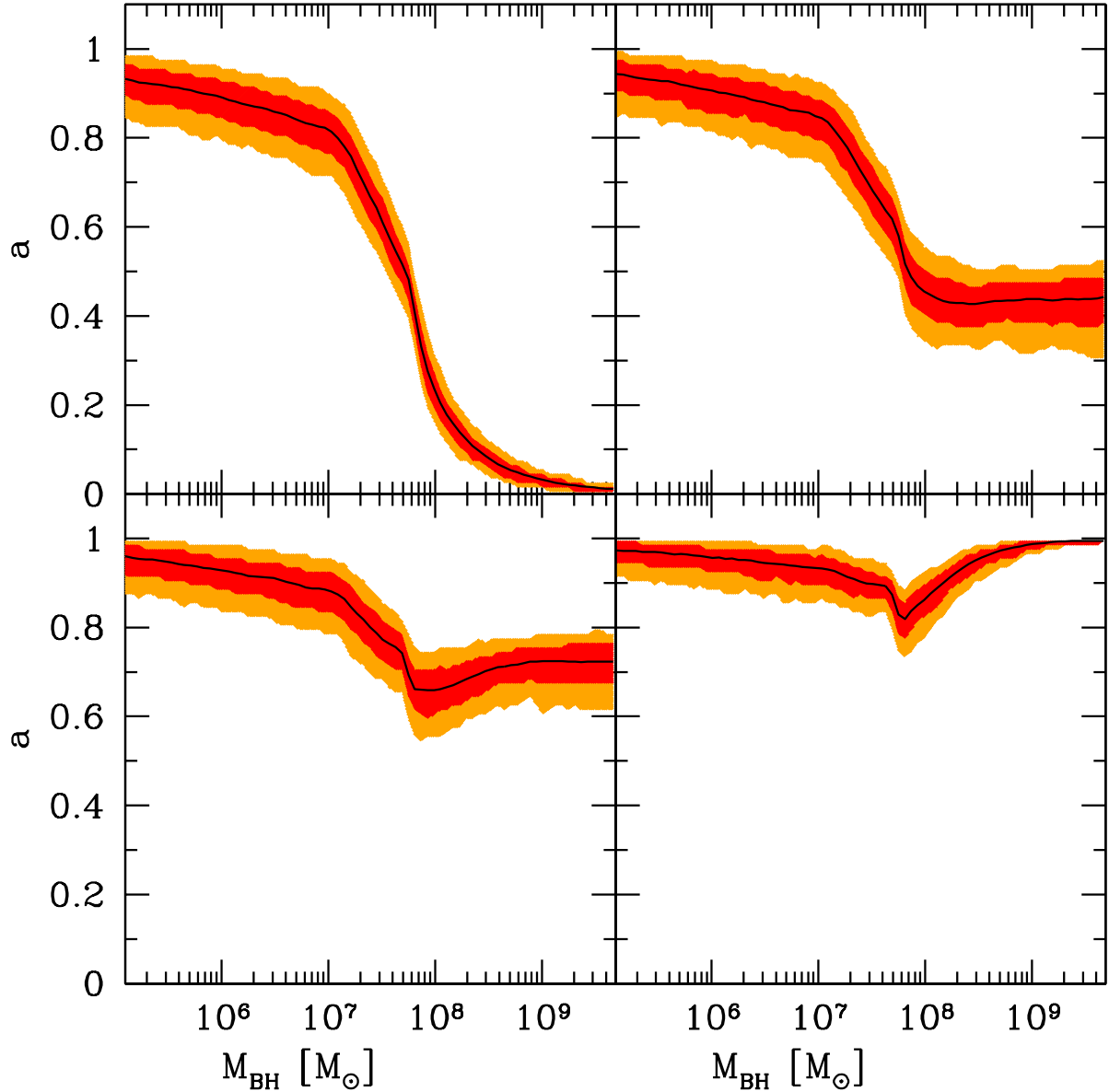}
\caption{Evolution of the BH spin magnitude $a$ as a function of
  $M_{\rm BH}$, assuming $m_{\rm cloud} =
  10^5\msun$. The upper left, upper right, lower left, and lower right panels
  refer to $F=0.5$, $0.25$, 0.125, and 0, respectively. 
  The colour code is as in figure~\ref{fig:a_cloud}.}
\label{fig:a_f}
\end{figure*}

Figure~\ref{fig:a_f} shows the spin evolution for different value of
the anisotropy parameter $F=0.5$; 0.25; 0.125; 0 (with $m_{\rm
  cloud}=10^5\,\msun$).  The early phase (I) is similar in all the
cases. Rapid BH-disk alignment results in $a\approx 0.9$.  At the
transition between phase I and II, the spin always decays, because at
this transition the BH-disk alignment is only partial, and retrograde
accretion is possible for a significant fraction of the time that
elapses during each accretion event.  By contrast, the value of the BH
spin at large masses ($\mbh\gsim 10^9\,\msun$) is strongly dependent
on the degree of anisotropy.  For $F=0.25,$ large BHs carry spins with
$a\approx 0.45,$ and this asymptotic value of $a$ increases with
decreasing $F$.  For $F=0$, i.e. the highest degree of anisotropy
described here, $a\to 1$.  This is a consequence of the stability of
the spin direction (discussed in section~\ref{sec:spindirection}).
For highly anisotropic cases, the massive BHs tend to align their
spins with the average angular momentum of the gas reservoir,
resulting in an higher fraction of prograde accretion events.  Thus,
we can correlate the spin of the large BHs with the dynamics of the
fueling mechanism.

\begin{figure*}
\includegraphics[width=0.96\textwidth]{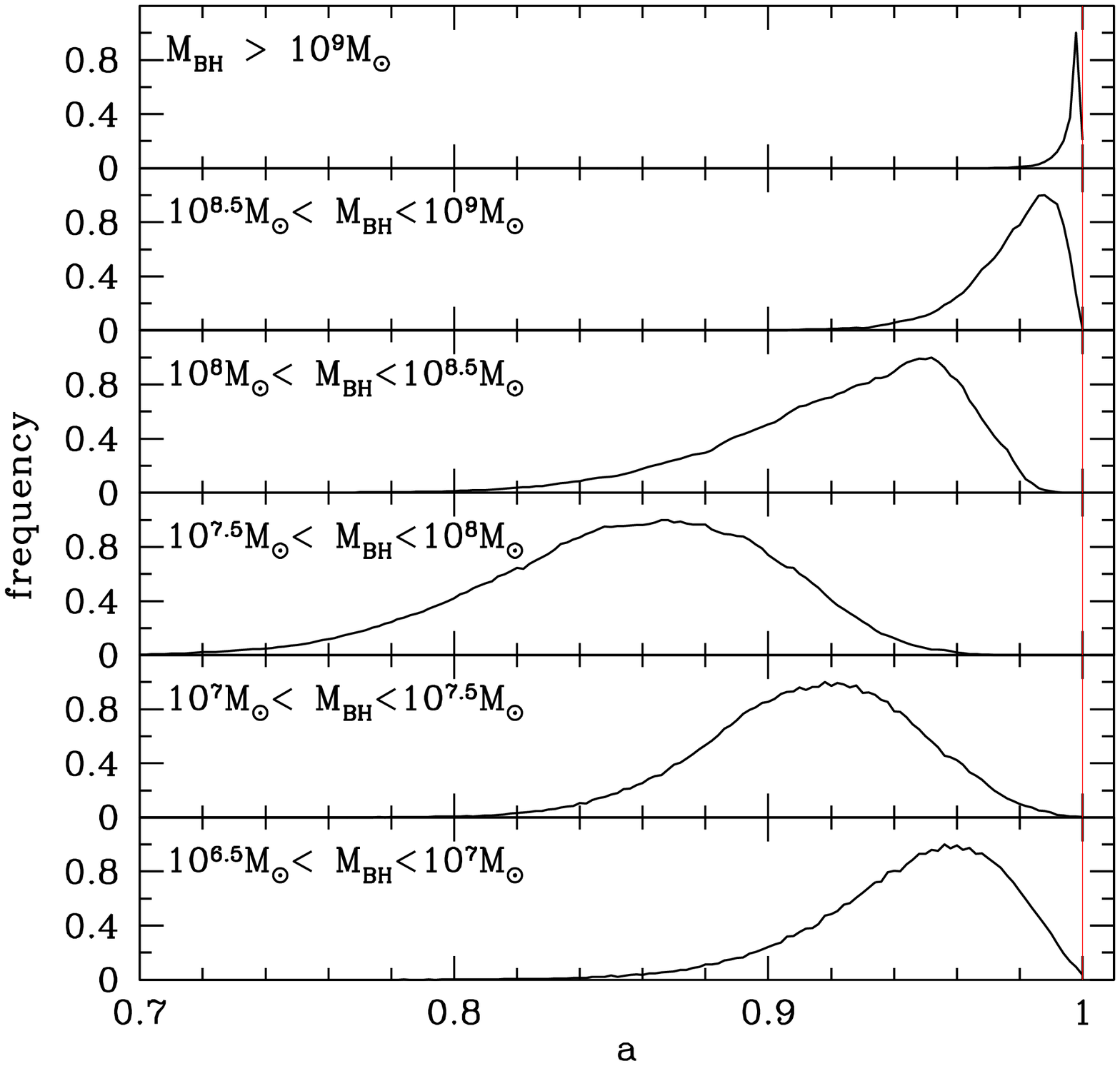}
\caption{Distribution of the BH spin magnitude $a$ for BHs in six
  different mass intervals, for $F=0$ and $m_{\rm cloud} =10^5
  \msun$. }
\label{fig:mass_a}
\end{figure*}

Figure~\ref{fig:mass_a} shows the distribution of the BH spin in
different mass intervals, assuming $F=0$ and $m_{\rm
  cloud}=10^5\,\msun$.  Here we emphasize that the highest spins are
attained at the largest BH masses.  The distribution corresponding to
$10^9\,\msun<\mbh<10^{9.5}\,\msun$ peaks at $a\approx 0.997$.  This
may have important implications on the efficiency at which these BHs
can accrete, on the statistics of very massive BHs, and on the
possibility of launching jets (e.g. Tchekhovskoy \& McKinney 2012).
A large degree of anisotropy ($F=0$) is the only way to
get maximally rotating BHs at the high mass end.

\section{Comparison with observations}

It is not straightforward to compare the predictions of our simple
investigation with constraints on the BH masses and spins inferred
from observations. Theoretically, a clear prediction of how the
expected spin distribution evolves with the BH mass can be made once
the distribution/dynamics of the fueling process is known (i.e., when
$F$ and $m_{\rm cloud}$ are known, in our context), and fueling can in
principle differ depending on the galaxy types and
masses. Observationally, the constraints on BH spins are often
debated. However, we can make some very general comments on how our
model fits within the observed trends.  As discussed in the
Introduction, the spin of a number of massive BHs has been measured
through spectral fitting of the broad K$\alpha$ iron lines at 6.4 keV.
In almost all the cases, the measured spins are larger than 0.5, for
BHs of masses $10^6 \, \msun \lsim \mbh \lsim 10^8 \,\msun$. The spin
parameters of the BHs hosted in Swift J2127, NGC 3783, and MGC-6-30-15
are still debated, and could be lower than 0.5, with a minimum of
$a<0.32$ for NGC 3783 (Patrick et al. 2011b, but see also Brenneman et
al. 2011 for a higher estimate of $a$). These measurements are in
agreement with our simple model, i. e.  there is always a history that
can recover the observed spin for the mass observed. In particular
these data suggest a non-negligible anisotropy in the fuelling of
these BHs. Note however that this could be related to an observational
bias toward measurements of highly spinning objects, as discussed in
Brenneman et al. (2011).

A possible evolution of the BH spin with mass
is found in the countinuity-equation based study by Shankar, Weinberg
\& Miralda-Escud\`e (2011). In this study, a radiative efficiency (and
as a consequence, a spin parameter) increasing with the BH mass is
needed to avoid overproduction of BHs at the high mass end of the
local observed mass function.  A similar study by Li, Wang \& Ho
(2012) finds $\eta$ increasing for increasing $M_{\rm BH}$ at high
redshift ($z \gsim 1$). We notice, however, that Li et al. do not
recover this trend at low redshift. They found the radiative
efficiency at $z \lsim 0.8$ to be almost independent on the BH mass.
The accretion efficiency could be derived fitting observational data
with accretion disk models for individual sources (Davis \& Laor 2011;
Raimundo et al. 2012). However, the values determined are largely
sensitive to the unknown parameters and uncertainties, which makes it
very hard to draw any conclusions on the spin dependence with black
hole mass, as discussed in Raimundo et al. (2012).
 
If the trend of $\eta$ suggested by Shankar et al. (and by Li et
al. for high redshifts) is confirmed, at least some massive BHs
($M_{\rm BH}\gsim 10^9 \msun$) must be very close to maximal rotation
($a \gsim 0.98$), because of the very steep evolution of $\eta$ close
to $a\approx 1$.  The dependence of $\eta$ on $a$ is quite sensitive
as $\eta$ varies from 0.151 for $a=0.90$, to $0.43$ for $a=1$.  BHs
can be spun up to such high spins ($a \gsim 0.98$) through BH-BH
binary coalescences only when specific conditions on the spins of the
parent BHs and the orbital configuration are met.  A coalescing binary
can result in a remnant BH with $a\approx 1$ only if $i$ the parent
BHs were close to maximal rotation prior to merge, and had their spins
aligned with the binary orbital angular momentum (Marronetti et
al. 2008; Kesden et al. 2010a)\footnote{This configuration is
  predicted for gas rich galaxy mergers (Bogdanovic, Reynolds \&
  Miller 2007; Dotti et al. 2010; Kesden et al. 2010b; Berti, Kesden
  \& Sperhake 2012).}, $ii$ through a long sequence of extremely
unequal mergers all confined in the same well defined orbital plane
(cf the "equatorial case" in Berti \& Volonteri 2008). For coalescing
BHs not rapidly rotating or with spins not aligned with their orbital
angular momentum, the remnant is expected to have a smaller spin, with
an average of $a \sim 0.7$ (Hughes \& Blandford 2003; Berti \&
Volonteri 2008).  To summarize, BH coalescences can preserve high
spins, but cannot be the original cause of extreme spins (Berti \&
Volonteri 2008).  The reason for close-to-maximally rotating BHs must
lie in their accretion history.  The very large spins in the most
massive BHs, if confirmed, will be an indication of a net degree of
anisotropy in the fueling of this objects ($F \approx 0$). We note
that our analysis predicts for the first time that the most rapidly
rotating BHs ($a > 0.98$) are also the most massive ones, and in
addition, that the spin orientation is stable over many accretion
cycles.

An independent constraint comes from jet observations. The bulk of
luminous radio-loud AGN is associated with very massive BHs
(e.g. McLure \& Jarvis 2004; Metcalf \& Magliocchetti 2006; Shankar et
al. 2008a,b; Shankar et al. 2010) typically hosted in very massive
ellipticals (e.g. Capetti \& Balmaverde 2006). If the radio jet is
powered by the extraction of rotational energy of the BHs, powerful
radio sources would be related to rapidly rotating BHs (e.g. Blandford
\& Znajek 1977). Recent MHD simulations suggest that the power is a
steep increasing function of the BH spin for $a\sim 1$
(e.g. Tchekhovskoy \& McKinney 2012, and references therein). In this
scenario, our model predicts the most massive BHs to produce the most
powerful jets, and jets are stable in their orientation if the BHs are
not completely isotropically fueled (but not necessarily experience
coherent accretion). The high mass BHs that have been fed by almost
isotropically distributed gas have lower spins, and lower radio
power. The difference in the fuelling process could explain the
observed dichotomy of the radio-loudnesses of AGN (Sikora, Stawarz \&
Lasota 2007). The prediction of the most powerful jets being related
to the most massive BHs is also in agreement with the observed trend
of the increasing fraction of radio-loud AGN for increasing BH masses
(McLure et al. 1999; Jiang et al. 2007; Caccianiga et al. 2010;
Chiaberge \& Marconi 2011), and with the evidence that the most
powerful radio and $\gamma-$loud AGN are associated to the heaviest
supermassive BHs in the universe, with masses $> 10^9\,\msun$
(Ghisellini et al.  2010a,b).

We can also compare the distributions of the spin orientations for BHs
of different masses as predicted by our model, with the observational
constraints available to date. It is commonly accepted that the
directions of relativistic jets in radio-loud AGN are not in a close
correlation with the galaxy morphology (e.g. Schmitt et al. 2002;
Verdoes Kleijn \& de Zeeuw 2005). However, different degrees of
alignment are observed, depending on the morphologies (and masses) of
the host galaxies.  Regarding Seyfert galaxies, the relative
orientation of the jets and the minor axes of the galaxy is consistent
with being drawn from an isotropic distribution (Clarke, Kinney \&
Pringle 1998; Nagar \& Wilson 1999; Kinney et al. 2000). The degree of
alignment seem to increase moving toward more massive galaxies and
more massive BHs. Schmitt et al. (2002) studied the degree of
misalignment between the direction of jets in radio-galaxies (early
type and S0) and the direction normal to their nuclear disks, as
traced by dust lanes. They found that the jets are typically
misaligned with respect to the normal to the nuclear disks by less
than $55^{\circ} \sim 80^{\circ}$. Similar results have been obtained
by Verdoes Kleijn \& de Zeeuw (2005) for low-power radio
galaxies. They found slightly better alignments when comparing the
direction of the jet and the normal to ellipsoidal-shaped dust lanes
(tracers of nuclear disks), with a peak in the distribution of
misalignments at $\sim 45^{\circ}$.  Such a level of alignment is in
contrast with an isotropic distribution, where a peak of misalignments
is expected to coincide with the peak of the solid angle, at $\sim
90^{\circ}$. Note however that, given the uncertainties in the
estimates, an isotropic distribution cannot be ruled out at more than
the $95\%$ confidence level (Verdoes Kleijn \& de Zeeuw 2005).
Verdoes Kleijn \& de Zeeuw suggest that while dust ellipses trace gas
on unperturbed Keplerian orbits that does not contribute to the
fueling of the central AGN, dust lanes can better trace perturbed gas
that traces the material falling toward the BH. The degree of
alignment between the normal to dust lanes and the jet direction is
sensibly higher, with a peak of the distribution at $20^{\circ} \sim
30^{\circ}$. These studies suggest that a relation between jets and
morphological properties of the galaxies exists only for massive
galaxies, hosting massive BHs. This result is in broad agreement with
our model (see Figure~\ref{fig:theta_distribution}) that also
indicates the stability of the jet direction, stability that is
necessary to launch a large-scale jet, as seen in radio galaxies.  We
notice that on average the direction of the jet is expected to be
related to the morphology of the nuclear distribution of gas, and not
necessarily to the large scale stellar distribution. However,
Saripalli \& Subrahmanyan (2009) found that the largest radio sources
(with a projected linear size exceeding 700 kpc) have the jets
preferentially aligned with the minor axes of their host galaxies. In
the framework of our model, such a stable direction of very large
scale jets can be explained if the BH is sufficiently massive ($\gsim
10^9 \msun$), and hosted in a galaxy with a significant degree of
anisotropy in its gas component. This trend is observed also by Browne
\& Battye (2010), as they observe ellipticals with a net rotation
hosting jets preferentially aligned with the rotation axes, while
non-rotating/triaxial ellipticals do not show such an alignment.

The predicted dependence of the magnitude and direction of $\rm{ J_{\rm BH}}$
on $F$ can be further tested studying the nuclear gas dynamics around BHs with
either an estimate of the spin parameter or a well constrained jet
direction. Owing to its extreme angular resolution, the Atacama Large
Millimeter/submillimeter Array is the
likely to be instrumental in
constraining the dynamics of dense molecular gas (and consequently $F$) in the
nuclei of nearby galaxies.

\section{Summary and discussion}
In this paper we explored the evolution of the spin parameter $a$ (or
$\rm{ J_{\rm BH}}$) of massive BHs considering the contemporary
evolution of the orientation $\jbhdir$, along sequences (histories) of
accretion episodes.  Histories are modeled as a succession of single
accretion events where an $\alpha-$disk of given mass $m_{\rm disk}$
and orientation $\jdisk$ forms.  Disk orientation is not fixed but it
is drawn from a distribution that carries a degree of anisotropy. This
anisotropy is expected to be seeded in the gas clouds that surround
the massive BH in the galactic nucleus and that are accreted.  Our
findings  and their astrophysical consequences can be summarized as follows:

\noindent 
$\bullet$ 
At  low BH masses ($\mbh\lsim 10^7\,\msun$)
the BH spin always aligns to the disk angular momentum in every single
accretion episode, and can be substantially misaligned relatively to
the average of the angular momenta of the disks.  For $\mbh\gsim
10^7\,\msun$, a single accretion episode does not modify significantly
the BH spin direction. In this regime, the BH spin aligns with the
average direction of the angular momentum of the accreting material,
with a degree of alignment that increases with increasing anisotropy
in the fueling process.

\noindent 
$\bullet$ {\it Small} BHs ($\mbh\lsim 10^7\,\msun$) carry {\it large}
spins $a\sim 0.9$. The {\it spin direction} changes {\it erratically}
from episode to episode so that the vector $\jbh$ exhibits a random
walk behavior, regardless the properties of the fueling process.

\noindent
$\bullet$ {\it Large} BHs ($\mbh\sim 10^9\,\msun$) can carry either
low spins or large spins, depending on the fueling conditions.  The
spin is low, $a\approx 0$, if the distribution of clouds in the hosts
is {\it completely} random and isotropic. The spin is larger, up to
$a\gsim 0.99$, and the {\it spin orientation} is {\it stable} on the
sky {\it if} the gas accreting onto the central BH posses some degree
of anisotropy, i.e. {\it if} the accreting material has, on average,
non zero angular momentum.  Only if the degree of anisotropy is high
($F=0$) the most massive black holes can be maximally rotating.

\noindent
$\bullet$ The most anisotropic case ($F=0$) studied here does not
correspond to what is usually referred to as coherent accretion: for
$F=0$, two successive accretion events can be misaligned by up to 180
degrees.  $F=0$ mimics accretion through disk clouds that are
distributed isotropically but that share a common sense of rotation.
$F=0$ corresponds to a 3D-dispersion to rotation velocity ratio
$\sigma/v_{\rm rot}\lsim 1$ for the gas fuelling the BH.

\noindent
$\bullet$ There may exist an interval of BH masses, in between the two
populations described above, in which the spins can be far from either
zero or unity and changes orientation at random.

\noindent 
$\bullet$  Although our analysis predicts that most low mass BHs have
substantial spins ($a\sim 0.9$; see the two bottom panels of
Figure~\ref{fig:mass_a}), the most rapidly rotating BHs ($a > 0.98$) are
also the most massive ones (see upper panel of Figure~\ref{fig:mass_a}).
In addition, the spin orientation of the most massive BHs remains
stable over many accretion cycles.

\noindent 
$\bullet$ The very large spins in the most massive BHs, if confirmed,
will be an indication of a net degree of anisotropy in the fueling of
this objects ($F \approx 0$). In other words, our model predicts that
a clear correlation exists between the kinematics of clouds feeding a
BH and its spin, but only in the case of the most massive BHs
($\mbh\gsim 10^7\,\msun$).

As a final comment, we notice that light BHs ($M_{\rm BH}\lsim
10^7\,\msun$) carry large spins undergoing erratic changes in their
orientation.  This can have strong implications on the efficiency of
feedback exerted by active BHs onto their host galaxy, that can
potentially set the BH-host galaxy scale relations (see Nayakshin,
Power \& King 2012 and references therein). The higher "inertia" of
the most massive BHs, if embedded in anisotropically moving gas, could
reduce the feedback efficiency, since only a small solid angle would
be affected.  By contrast, if any anisotropic feedback (such as jets
or biconical outflows, etc.) are launched during single episodes
around less massive BHs, the spin random walk would result in a spread
of the injected energy over $4 \pi$ during their lifetime.

Future studies will address the consequences of the model in the
context of the cosmological evolution of BHs.

\section*{Acknowledgments}

We thanks the anonymous Referee, Enrico Barausse, Alessandro
Caccianiga, Francesco Haardt, Sandra Raimundo, Alberto Sesana,
Francesco Shankar for comments and suggestions. MV acknowledges funding
support from NASA, through award ATP NNX10AC84G; from SAO, through
award TM1-12007X, and from a Marie Curie Career Integration grant
(PCIG10-GA-2011-303609).

\label{lastpage}

\end{document}